\documentclass[aps,prl,twocolumn,showpacs,superscriptaddress]{revtex4}
\usepackage{graphicx}
\usepackage{dcolumn}
\usepackage{bm}
\usepackage[final]{pdfpages}
\usepackage{graphicx}

\begin{document}

\title{The electronic transport of top subband and disordered sea in InAs nanowire in presence of a mobile gate.
}

\author{A.A.~Zhukov}
\affiliation{Institute of Solid State Physics, Russian Academy of
Science, Chernogolovka, 142432 Russia}

\author {Ch.~Volk}
\author {A.~Winden}
\author {H.~Hardtdegen}
\affiliation{Peter Gr\"unberg Institut (PGI-9), Forschungszentrum
J\"ulich, 52425 J\"ulich, Germany} \affiliation{ JARA-Fundamentals
of Future Information Technology, Forschungszentrum J\"ulich,
52425 J\"ulich, Germany}
\author {Th.~Sch\"apers}
\affiliation{Peter Gr\"unberg Institut (PGI-9), Forschungszentrum
J\"ulich, 52425 J\"ulich, Germany} \affiliation{ JARA-Fundamentals
of Future Information Technology, Forschungszentrum J\"ulich,
52425 J\"ulich, Germany}
\date{\today}

\begin{abstract}
We performed measurements at helium temperatures of the electronic
transport in an InAs quantum wire ($R_{wire} \sim 30$\,k$\Omega$) in the presence of a charged tip of an atomic force microscope serving as
a mobile gate. The period and the amplitude of the observed
quasiperiodic oscillations are investigated in detail as a
function of electron concentration in the linear and non-linear
regime. We demonstrate the influence of the tip-to-sample distance on the ability to  locally affect the top subband electrons as well as the electrons in the disordered sea. Furthermore, we introduce a new method of detection of the subband occupation in an InAs wire, which allows us to evaluate the number of the electrons in the conductive band of the wire.
\end{abstract} 

\maketitle

\section{Introduction}

In the past decade an increasing number of investigations were
dedicated to the electronic transport in semiconductor 
nanowires \cite{Li2006,Thelander2006,Appenzeller2008,Yang2010}. Especially nanowires based on III-V semiconductors, e.g. InAs,  are very attractive as conductive channels in devices for nano-electronic applications \cite{Bryllert2006,Dayeh2007}. Apart from more application-driven investigations, InAs nanowires are also very suitable objects to study fundamental quantum phenomena, i.e. single electron tunneling \cite{Fasth2005,Shorubalko2007} or electron interference \cite{Dhara2009,Roulleau2010,Hernandez2010,Bloemers2011}, at low temperatures. At helium temperatures the transport in InAs nanowires is mostly diffusive and typical values of the elastic mean free path $l_e$ are of the order of a few tenth of nanometers \cite{Dhara2009,Roulleau2010,Bloemers2011}. Information on electron phase coherence can be extracted from the temperature dependence of universal conductance fluctuations \cite{Hernandez2010,Bloemers2011}.

In order to gain detailed information of local conductance features in
low-dimensional systems, mobile gate measurements employing a charged AFM tip (scanning gate microscopy measurements or SGM measurements) have been established as a standard method. Quantum point contacts
\cite{TopinkaS2000,TopinkaN2001,SchnezQPC2011}, quantum rings \cite{Hackens2006}, quantum dots based on heterojunctions \cite{Pioda2004,Gildemeister2007,Huefner2011},
graphene \cite{Schnez2010}, and carbon nanotubes
\cite{Bockrat2001,Woodside2002,Zhukov2009} have been
investigated comprehensively using SGM. Investigations of local electronic transport in InAs nanowires with scanning gate microscopy were performed at room temperature \cite{Zhukov2011E,Zhou2007} as well as at He temperatures
\cite{Bleszynski2005,Boyd2011a,Zhukov2011,Zhukov2012,Zhukov2012P,Zhukov2013}.
However, these studies mostly focused on wires with initially
existing \cite{Bleszynski2005,Zhukov2011,Zhukov2012,Zhukov2013} or
artificially created \cite{Boyd2011a} defects, i.e. potential barriers
made of InP, which divided the nanowire into series of quantum
dots.

Recently, in InAs nanowires without defects with characteristic
resistance values of 30\,k$\Omega$ unexpected quasi-periodic
oscillations of the resistance along the wire were observed in SGM
scans \cite{Zhukov2012P}. The non-monotonic dependence of the
period of the observed  oscillations on the back-gate voltage
allowed to associate them with electrons in the top subband with a
small Fermi wavelength comparable with the length of the wire
($\lambda_F \sim l_{wire}$) altering the resistance of the whole
system. These electrons do not scatter on the surface of the
nanowire and do not mix with other free electrons of the lower
laying mixed subbands (disordered sea). However, the experiment
reported in Ref.~\cite{Zhukov2012P} left the mechanism behind the
oscillations as an open question.

Here, we present a detailed investigation of the resistance oscillations in an InAs nanowire, namely the tracing of the oscillation from a clearly defined two-nodes state through a three-nodes one to a state where minima split. Furthermore the stability of two-nodes
oscillations in the non-linear regime is discussed. We demonstrate the
influence of the tip-to-sample distance on a ability to
locally influence the top subband electrons as well as the
electrons in the disordered sea. We suggest a new method to
determine the occupation of the topmost subband using a line trace
of a charged AFM tip along the wire. This method allows us to
evaluate the number of conductive electrons added to the InAs wire
on applying a positive back-gate voltage.

\section{Experimental}

In our experiment we study a nominally undoped InAs nanowire grown
by selective-area metal-organic vapor-phase epitaxy \cite{Akabori2009}. The diameter of the wire is 100\,nm. The wire was placed on an
$n$-type doped Si (100) substrate covered by a 100~nm thick
SiO$_2$ insulating layer. The Si substrate served as the back-gate
electrode. The evaporated Ti/Au contacts to the wire as well as
the markers of the search pattern were defined by electron-beam
lithography. The distance $l_{wire}$ between the contacts is
$2.6\,\mu$m. A scanning electron beam micrograph of the
sample is shown in Fig.~1a). The source and drain metallic
electrodes connected to the wire are marked by S and D .
\begin{figure}
\includegraphics*[width=0.8\columnwidth]{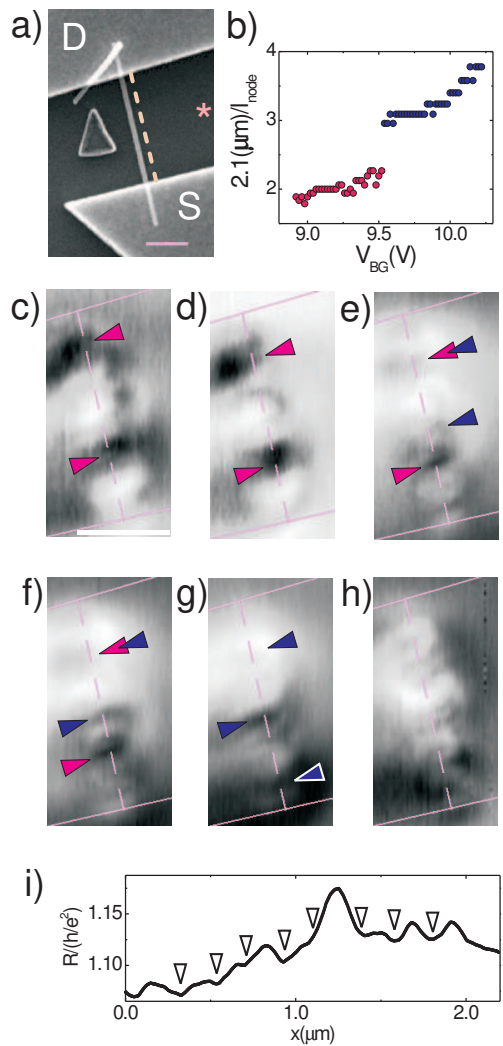}
\caption {(color online) a) Scanning electron microscope image of
the InAs nanowire. The source and drain contact pads are marked by S
and D. The scale bar corresponds to 1\,$\mu$m. The positions of
the tip during the transport measurements vs $V_{BG}$ performed
equidistantly along the wire are marked by a dashed line, the
position of tip for the reference measurement is marked by '$\star$'. b) Dependence of 2.1\,$\mu$m/$l_N$, where $l_N$ is the distance between nodes of the resonances extracted from the set of SGM images. Bright
and dark symbols show $l_{node}$ for $N_{node}=2$ and 3,
respectively. c) to h) SGM images measured at $V_{BG}=8.98$,
9.14, 9.50, 9.62, 9.94 and 10.40\,V, respectively, depicting the
nanowire resistance at different tip positions ($V_{tip}=0$\,V).
Brighter color means higher resistance. The driving current is $I_{AC}=1$\,nA. The tip to surface distance is $h_{tip}=300$\,nm. The positions of the nodes are marked with pink and blue triangles for two-nodes and three-nodes patterns, respectively. The horizontal scale bar in c) corresponds to 1\,$\mu$m. The scale is the same for all
SGM images from c) to h). The solid lines in each SGM image traces
the edges of the metallic contacts while the dashed line marks the
wire axis. i) Crosscut of h) along the wire axis
with minima marked by triangles.} \label{Fig1}
\end{figure}

All measurements were performed at $T=\,4.2$\,K. The charged tip of a
home-built scanning probe microscope \cite{AFM} is used as a
mobile gate during scanning gate imaging measurements. All
scanning gate measurements are performed by keeping the potential of
the scanning probe microscope tip ($V_{tip}$) as well as the back-gate
voltage ($V_{BG}$) constant. The conductance of the wire during
the scan is measured in a two-terminal circuit by using a standard
lock-in technique. Here, a driving AC current with an amplitude of
$I_{AC}=1$\,nA at a frequency of 231\,Hz is applied while the voltage is measured by a differential amplifier. Two typical tip to SiO$_2$ surface distances were chosen for the scanning process $h_{tip}=300$\,nm, i.e. the tip is far from the surface and $220$\,nm, i.e. the tip is close to the surface.

\section{Experimental results}

In Figs.~1c) to h) scanning gate microscopy images in the linear regime are shown which are obtained at back-gate voltages of $V_{BG}=8.98$, 9.14, 9.50, 9.62, 9.94 and 10.40~V, respectively, keeping the tip voltage $V_{tip}$ fixed at zero. The tip-to-surface distance during these measurements was kept at $h_{tip}=300$\,nm. Increasing the
back-gate voltage the situation develops through a well-defined
two-nodes pattern [cf. Figs.~1c) and d)], then passing a
two-to-three nodes regime [cf. Figs.~1e) and f)] and subsequently
to a well-defined three-nodes one [cf. Fig. 1g)]. The corresponding positions of nodes in the SGM scans are marked by pink and blue triangles for two-nodes and three-nodes patterns, respectively.
At $V_{BG}=9.14$ and 9.94~V in the \emph{single resonance regime} [cf. Figs. 1c) and g)] the oscillations are well-defined, while in the intermediate regime the oscillations are slightly less visible [cf. Figs. 1e) and f)]. Finally, at the highest back-gate voltage of 10.40~V, the pattern evolves into a four-nodes regime with split minima, as can be inferred from Fig.~1h). A crosscut of this pattern along the wire axis is shown in Fig.~1i). The resulting eight quasi-periodic minima are marked by down-nose triangles.

In order to access the rigidity of the observed oscillations we calculated the ratio $l_{eff}/l_{node}$, with $l_{eff}$ the effective wire length and $l_{node}$ the distance between the nodes. As can be seen in Fig.~1b), the ratio defined above strongly depends on the back-gate voltage $V_{BG}$ and resembles a staircase shape with well developed plateaus. In order to adjust the position of the lower step to an integer node number $N_{node}$, we used a value of $2.1\,\mu$m for the effective wire length $l_{eff}$. This value is slightly smaller than the geometrical length of the wire $l_{wire}=2.6\,\mu$m but looks reasonable keeping in mind the presence of the depletion regions at the interfaces between the wire and the metallic contacts.

Next we present scanning gate measurements, where the driving current through the nanowire was increased stepwise. The images shown in Figs.~2a) to d) correspond to a $I_{AC}=1$, 4, 12.5 and 50\,nA, respectively. These measurements were obtained by applying a
back-gate voltage of $V_{BG}=8.7$\,V and setting $V_{tip}$ to zero. 
Once again the tip is kept far from the surface ($h_{tip}=300$\,nm). 
It is worth noting that while in Figs.~2a) to c) the energy corresponding to the typical source-to-drain voltage is smaller than the thermal energy $eV_{SD} \lesssim k_BT$, while for the case of Fig.~2d) the situation is opposite i.e. $eV_{SD} \sim 1.6$\,meV$ \gg k_BT$, with $k_B$ is the Boltzmann constant. Crosscuts of Figs.~2a) to d) along wire axis are shown in Fig.~2e). As can be seen here, no significant deviations of the node positions and the amplitude of the oscillations are found up to $I_{AC}= 12.5$\,nA. However, a slight suppression of the oscillations and a decrease of the resistance are observed in the strong non-linear regime at $I_{AC}=50$\,nA.
\begin{figure}
\includegraphics*[width=0.8\columnwidth]{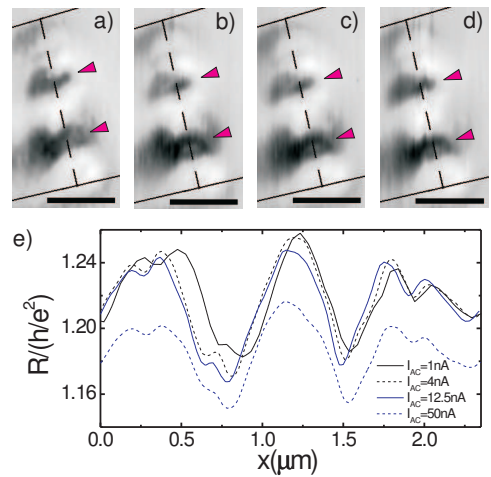}
\caption{ (color online) a) to d) SGM images of the InAs nanowire
resistance (measured source to drain voltage divided by driving
current) gained by using a driving current of $I_{AC}=\,1$, 4,
12.5 and 50\,nA, respectively. The back-gate voltage of
$V_{BG}=8.7$~V and tip voltage of $V_{tip}=0$~V are kept at the
same values for all four scans. The tip-to-surface distance is
$h_{tip}=300$\,nm. The positions of the nodes are marked by pink
triangles. Brighter colors mean higher resistances. The
horizontal scale bar in each SGM scans corresponds to 1\,$\mu$m.
The solid lines in each SGM image trace the edges of metallic
contacts while the dashed line marks the wire axis. e) Crosscuts
of SGM scans shown in a) to d) normalized to the quantum resistance
($h/e^2$). Solid and dashed black lines correspond to driving
currents of $I_{AC}=$\,1 and 4~nA, solid and dashed blue lines
correspond to driving currents of $I_{AC}=$\,12.5 and 50\,nA. }
\label{Fig2}
\end{figure}

For the next two sets of SGM images shown in Figs.~3a) to f) a closer  tip-to-surface distance of $h_{tip}=\,220$\,nm was chosen. These measurements were performed in the linear regime by using a driving current of 1\,nA. The first set, shown in Figs. 3a) to c), is measured at smaller back-gate voltages of $V_{BG}=\,3.97$, 3.98 and 3.99\,V
The triangles mark the nodes of three-nodes pattern created by conductive electrons of the top subband. The second set of scans depicted Figs.~3d) to f) is taken at higher back-gate voltage of 8.51, 8.52 and 8.53\,V, respectively. Here, the smooth background was subtracted, in order to emphasize the small scale ripples on the scan. As a reference, Fig.~3g) shows crosscuts along the wire axis of pristine SGM data (without subtraction) at the different back gate voltages.
\begin{figure}
\includegraphics*[width=0.8\columnwidth]{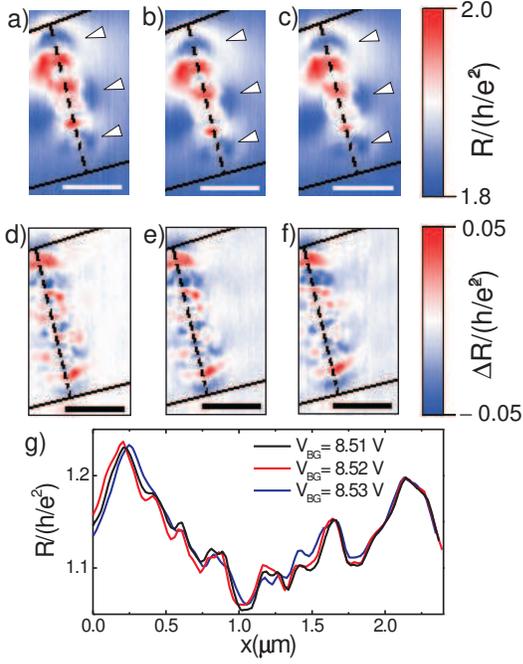}
\caption{a) to c) SGM images of the InAs wire resistance
made using driving currents of $I_{AC}=\,1$\,nA. A shorter tip-to-surface distance of $h_{tip}=\,220$~nm was chosen. The tip voltage is $V_{tip}=\,0$\,V for all scans. Back-gate voltages are of $V_{BG}=\,3.97$, 3.98 and 3.99~V for a) to c), respectively. The side scale bar represents the color to resistance mapping in quantum resistance units ($h/e^2$). Triangles mark nodes of three-nodes pattern. d) to e) are SGM images of the InAs nanowire resistance made using driving currents of $I_{AC}=\,1$\,nA after subtraction of the smooth background $\Delta R=R-R_{smooth}$ in quantum resistance units ($h/e^2$). Back-gate voltages are of $V_{BG}=\,8.51$, 8.52 and 8.53\,V for d) to f), respectively. The side scale bar represents the color to $\Delta R/(h/e^2)$ map. The horizontal scale bar in all SGM scans corresponds to 1~$\mu$m. The solid lines in each SGM image traces the edges of the metallic contacts and the dashed line marks the wire axis. g) represents crosscuts of SGM scans d) to f) normalized on quantum resistance ($h/e^2$) before background subtraction.} \label{Fig3}
\end{figure}

The next set of experimental data is dedicated to resistance vs.
back-gate dependencies for different charged tip positions. Once again, for all measurements the tip voltage is set to zero. First, in  Fig.~4a) the dependence of the normalized reciprocal wire resistance $(h/e^2)/R_{ref}$, i.e. the normalized conductance, on the back-gate voltage $V_{BG}$ is shown,  where the tip is placed a a fixed location far from the wire. The location of the spot is indicated by a $\star$ in Fig. 1a). The trace shown in Fig.~4a) will serve as a reference for the later position dependent measurements. The fluctuating conductance can be assigned to the phenomena of universal conductance fluctuations \cite{Altshuler1985b,Lee1987}. The corresponding fluctuation amplitude can be determined by first subtracting the background conductance $(h/e^2)/R_{lin}(V_{BG})$ obtained from a linear fit of $(h/e^2)/R_{ref}(V_{BG})$. The remaining fluctuations $\Delta((h/e^2)/R)=(h/e^2)(1/R_{ref}-1/R_{lin})$ are shown in 
the inset of the Fig.~4a). Here, a typical value
of fluctuation amplitude of $0.1(h/e^2)$ for the back-gate voltage range of $0 \leq V_{BG} \leq 12$\,V. This value fits well to comparable measurements on InAs nanowires \cite{Hernandez2010,Bloemers2011}.

Next we will present a set of measurements, where the tip is placed equidistantly from source to drain at fixed positions $i=1 \ldots 11$ along the dashed line indicated in Fig.~1a). In Fig.~4b) the deviation of the conductance with respect to the conductance with a tip positions far from the wire $R_i^{-1}(V_{BG})-R^{-1}_{ref}(V_{BG})$ is plotted as a function of back-gate voltage for the different positions $i$.
Three well defined minima in the conductivity (grooves) are
observed around $V_{BG} \approx 2.7$, 7.5 and 10.5\,V marked by I,
II and III in Fig.~4b).
\begin{figure}
\includegraphics*[width=0.8\columnwidth]{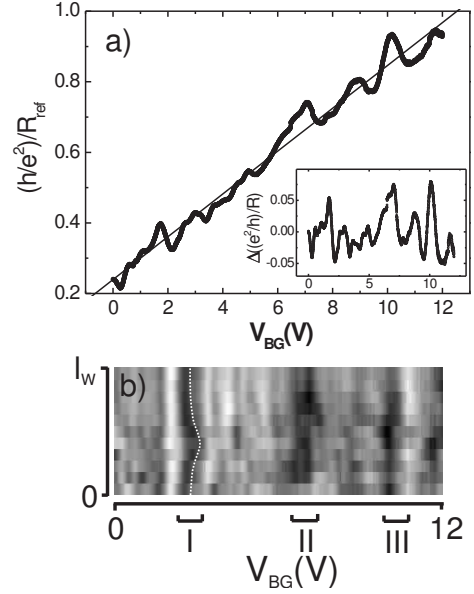}
\caption{(color online) a) $(h/e^2)/R_{ref}$ of the wire vs
back-gate voltage measured when the tip ($V_{tip}=0$~V) is placed at
position of (*) (see Fig. 1a)). The solid line is the linear fit
$1/R_{lin}$ and the inset is the dependence
$\Delta((h/e^2)/R)=(h/e^2)(1/R_{ref}-1/R_{lin})$. b) Mapping of
$1/R(i)-1/R_{ref}$ as the function of $V_{BG}$ and the tip
position along the wire (cf. Fig. 1a)) $i$ ($i=1..11$). Light gray
correspond to positive values while dark grey means negative
values. The three voltage ranges where the new subband arises in
the wire I ($V_{BG} \approx 2.7$~V), II ($V_{BG} \approx$7~V) and
III ($V_{BG}\approx$10.5~V) are marked (see the text for details).
The white dotted line is a guide to the eyes and generally shows
the profile of the groove at $V_{BG}\approx 2.7$~V. } \label{Fig4}
\end{figure}

\section{Discussion}

In our previous work SGM experimental data with 10, 2, 3 and 4
standing wave nodes ($N_{node}$) have been presented
\cite{Zhukov2012P}. These experimental results were obtained in
the previously mentioned resonance condition
$N_{node}l_{node}=l_{wire}$. While the nodes are the most
pronounced features, no information about $l_{node}(V_{BG})$ in
between resonances were provided. It was found in Ref.~\cite{Zhukov2012P}, that the oscillation amplitude
drops as $N_{node}$ is increasing, thus the transformation from
$N_{node}=2$ to $3$ was chosen in the present study as the
preferable $V_{BG}$ range to extract $l_{node}(V_{BG})$ in out of
resonance condition. The observed staircase shape dependence of
$1/l_{node}$ vs. $V_{BG}$, as indicated in Fig.~1b), is
qualitatively in good agreement with the model of standing waves
of ballistic electrons in the top transverse quantization subband.
Standing waves are the result of the presence of potential
barriers at metal-semiconductor interfaces. In the models of Wigner
crystallization  of Luttinger liquid and Friedel oscillations a
similar dependence of $1/l_{node}$ vs. $V_{BG}$ is expected
\cite{Ziani2012}. It is worth noting that the barrier at the top wire
to contact interface is more pronounced according to SGM pictures,
so standing wave appears to be ``pinned'' to the top contact, i.e.
the top contact to node distance increases as $V_{BG}$ increases.

One finds, that the kinetic energy
$E_k(N_{node}=3)=[(3\pi/l_{wire})\hbar]^2/(2m^*)=21\,\mu$eV is
essentially less than $k_BT$ even for the three-nodes resonant
state. Here, $m^*=0.023\,m_e$ is the effective mass of electron in
InAs, $m_e$ is the free electron mass, $\hbar$ is the Planck
constant. In addition, the thermal length
$L_T(N_{node}=3)=h^2/(2\pi m^*k_BT(2L/3))\sim 30$\,nm is
considerably smaller than the length of the wire
$l_{wire}=2.6$\,$\mu$m. The energy difference between the two-nodes
and three-nodes resonance states is less than $k_BT$ because both
states are resolved simultaneously at a back-gate voltage
$V_{BG}=9.62$~V [cf. Fig.~1f)]. The Coulomb energy of the
electron-electron interaction,
$V_C(N_{node}=3)=2e^2d_d^2/(4\pi\epsilon_0\epsilon)(l_{wire}/3)$,
even for the three-nodes resonant state is less than $k_BT$ as
well. Here, $\epsilon=15.15$ is the static dielectric constant in
InAs, $d_d \sim 150$~nm is the distance from the center of the
wire to the doped back-gate, and $\epsilon_0$ is the vacuum
permittivity. Thus, the ability of Wigner crystallization,
Friedel oscillations and standing wave scenarios looks rather
doubtful for $N_{node}=2$ and 3. The Coulomb interaction between
electrons becomes comparable to the temperature $E_C \sim
0.16$\,meV only at $V_{BG}=10.40$\,V when the average distance
between the electrons is around 250\,nm. We may speculate that the
formation of a Wigner crystal in the top subband happens at this
back-gate voltage with a $2k_F$ to $4k_F$ transformation of the
oscillations.

Nevertheless, the experimentally observed resistance oscillations
are apparently connected to the formation of charge density
modulations in the top subband. Besides this, the mechanism of the
effect of the charge density oscillations on the resistance of the
whole wire is more due to the tip-induced spatial fluctuations of
the chemical potential of the top subband \cite{Ziani2012,Gindikin2007,Soeffing2009}. Additionally, some suppression of the conductance of the disordered sea due to the increasing charge density in the top subband in the node location similar to \cite{Tans,Zhukov2011E} is expected. Thus two mechanisms alter the resistance of the whole wire in opposite direction.

SGM images presented in Fig.~2 demonstrate the robustness of the
two-nodes resonant state to application of source to drain
voltage. As it was mentioned above, no significant deviations of
the node positions and the amplitude of the oscillations are found
as long as $eV_{SD}\le k_BT$. Only the application of a large
current of $I_{AC}=50$~nA resulting in $eV_{SD}\sim
1.6\,$meV$\,\gg k_BT$ suppresses the oscillations and decreases
the resistance [cf. Fig.~2e)].

The results of the SGM scans made when the tip to surface distance
is 220\,nm, i.e. the closest distance from tip to the wire axis is
of 170~nm, are presented in Figs.~3a) to g). As previously stated,
if the distance from the tip to the wire axis $l_{tip}\gg d$ the
density oscillations of the top sublevel are probed while at
$l_{tip} \sim d$ the charged tip disturbs the disordered sea as
well. The characteristic deviation of the back-gate voltage
inducing the redistribution of conductivity maxima and minima in
SGM images when tip is scanning over the nanowire ($l_{tip} \sim d$)
is about 10\,mV. In contrast to that, the width of the
step in dependence of $l_{node}(V_{BG})$ is more than hundred
millivolts. The irregular pattern observed in SGM scans governed
by the disordered sea has the smallest typical length scale of
200~nm, this is in accordance with expected spatial resolution of
the experimental set up for $h_{tip}=220$~mn. The amplitude of the
deviations of resistance [cf. Fig. 3g)] is comparable with the
amplitude of the universal conductance fluctuations shown in Fig.
4a) (inset).

In Ref.~\cite{Zhukov2012P} the abrupt increasing of the $l_{node}$
was interpret as the formation of the new subband in the InAs
wire. Just at the instance the subband formation electrons are
loaded simultaneously to the disordered sea and to the band with
increasing of the back-gate voltage. Electrons loaded to the new
subband are blocked because of the potential barriers at the wire
to contact interfaces forming a semi-opened quantum dot. This dot
decreases the total conductance of the wire at certain range of
back-gate voltage, namely from $V_{BG}$ of quantum dot formation
to the value of the back-gate voltage when the barriers become
transparent. Lets call the center of this range $V_{BG1}$. It is
possible to slightly alter $V_{BG1}$ with the charged AFM tip.
Taking into account that electrons in quantum dot are concentrated
in the center of the wire, $V_{BG1}$ as a function of the tip
position along the wire must have one maximum
\cite{Zhukov2009,Boyd2011}. This maximum is traced by the dashed
line in Fig.~4b). Thus, three subbands are formed marked as I, II
and III in the region of back-gate voltage of
0\,V$\leq V_{BG}\leq 12$\,V at $V_{BG} \approx 2.7$, 7.5
and 10.5\,V, respectively. There is some discrepancy for back-gate
voltages less than 1~V in the determination of the formation of
the second subband comparing to the data in Ref.
\cite{Zhukov2012P}, where the formation was determined directly
from SGI scans. It originates from the hysteresis in
$R_{wire}(V_{BG})$ of the sample.

The formation of three subbands in the 0~V$<V_{BG}<12$~V back-gate
region means that the number of free electrons loaded into wire is
less than 100, while the total amount of electrons added to wire
calculated from the capacitance is  around 2000
\cite{Zhukov2012P}. It means that most of the electrons in the
wire are trapped probably because of interface states charged
by changing the back-gate voltage.
This also means that the value of mean free path
must be recalculated as well. It seems $l_e$ is actually larger
than the typically used value of 40~nm even for disordered sea. We
would like to remind that a value of $l_e \sim 200$~nm was
measured by Zhou {\it et al.} \cite{Zhou} at room temperature.

\section{Conclusion}

We performed measurements at helium temperatures of the electronic
transport in an InAs nanowire ($R_{wire} \sim 30\,$k$\Omega$) in
the presence of a charged tip of an atomic force microscope
serving as a mobile gate. The period and the amplitude of
previously observed quasi-periodic oscillations are investigated
in detail as a function of back-gate voltage in the linear and
non-linear regime. None of the scenario such as Friedel
oscillations, Wigner crystallization or standing wave looks
applicable to explain the origin of observed oscillations at
$N_{node}=2$ and 3. We demonstrate the influence of the
tip-to-sample distance on an ability to influence locally the
top subband electrons as well as the electrons in the disordered
sea. We suggest a new method of evaluation of the number of
conductive electrons in an InAs wire. This method results in the
conclusion that most of the electrons added to nanowire conductive
band on applying a positive back-gate voltage are trapped.

\section{Acknowledgments}

This work is supported by the Russian Foundation for Basic
Research, programs of the Russian Academy of Science, the Program
for Support of Leading Scientific Schools, and by the
International Bureau of the German Federal Ministry of Education
and Research within the project RUS 09/052.

{}


\end{document}